# A Finite-temperature First-principles Approach to Spin Fluctuation in BaFe$_2$As$_2$

Y. Wang, S. L. Shang, L. Q. Chen, and Z. K. Liu

Materials Science and Engineering, The Pennsylvania State University,

University Park, PA 16802, USA

Thermodynamic flunctuations in BaFe$_2$As$_2$ is addressed with a first-principles formulation of the Helmholtz energy by accounting for the mixture of various electronic states each distinguished by different spin orientation distributions. We find that it is the spin exchange coupling in the inter-plane $c$ direction that dictates the spin density wave ordering. We quantitatively predicted the pressure dependence of the spin density wave ordering, the Schottky anomaly, and the temperature dependence of thermal populations of spin structures, all in agreement with available experimental data.

The surprising discovery of the relatively high-temperature superconductivity in iron pnictides in 2008 [1] has it more enigmatic of the origin of high-temperature superconductivity. There is no widely accepted theory on the subject after over 100,000 published papers since the finding of a lanthanum-barium-copper oxide ceramic in 1986 [2]. Other examples include heavy Fermion systems [3] and ferromagetic superconductors [4, 5] whose superconductivity cannot be understood in the existing framework of BCS theory [6].

In this Letter, we develop a finite-temperature first-principles approach to reveal the spin fluctuation in $BaFe_2As_2$ [7-9] – one of the parent compound of the iron pnictide superconductors in which the electrons of the neighboring Fe atoms interact in a coordinated manner and behave collectively.

For a lattice with $N$ atoms where volume $V$ and temperature $T$ are constant, assuming fluctuated spin distributions, the partition function, $Z$, can be written as [10]:

$$Z = \sum_{\sigma} w^{\sigma} \sum_{i \in \sigma, \rho \in \sigma} \exp[-\beta \varepsilon_i(N,V,\rho)], \qquad (1)$$

where $\sigma$ labels the electronic state, each distinguished by unique spin orientation distributions in the lattice, $\beta = 1/k_B T$, $w^{\sigma}$ is the multiplicity of spin structure $\sigma$, $i$ identifies the vibrational states belonging to the spin structure $\sigma$, $\rho$ labels the electronic distributions associated with spin structure $\sigma$, and $\varepsilon_i(N,V,\rho)$ is the eigenvalue of the corresponding microscopic Hamiltonian. We first complete the summation over all states belonging to a specific spin structure $\sigma$. This gives

$$\sum_{i\in\sigma,\rho\in\sigma}\exp[-\beta\varepsilon_i(N,V,\rho)]=Z^\sigma=\exp[-\beta F^\sigma(N,V,T)]. \tag{2}$$

It is immediately apparent that $Z^\sigma$ is the partition function of spin structure $\sigma$, $x^\sigma = w^\sigma Z^\sigma / Z$ is its thermal population, and $F^\sigma(N, V, T)$ is its Helmholtz free energy. With $F = -k_B T \log Z$ [10], we find

$$F(N,V,T)=\sum_\sigma x^\sigma F^\sigma(N,V,T)-TS_f(N,V,T), \tag{3}$$

where the spin fluctuations results in the configurational entropy

$$S_f(N,V,T)=-k_B\sum_\sigma x^\sigma \log(x^\sigma/w^\sigma). \tag{4}$$

We calculate $F^\sigma(N,V,T)$ using [11]

$$F^\sigma(N,V,T)=E_c^\sigma(N,V)+F_v^\sigma(N,V,T)+F_{el}^\sigma(N,V,T). \tag{5}$$

Note that $E_c^\sigma$ is the 0 K static total energy, $F_v^\sigma$ is the vibrational free energy, and $F_{el}^\sigma$ is the thermal electronic free energy.

We treat the spin fluctuation in BaFe$_2$As$_2$ with a thermodynamic system of $\sqrt{2}a \times \sqrt{2}a \times 4c$ 40-atom supercell. Considering only the spin up and down flips of the 16 Fe atoms, such a system possesses 65536 spin configurations [12]. Out of the 65536 configurations, only 13 spin structures, which are built by modulating, along $c$ direction, the "2d-Stripe" structures [13-17], are needed to be considered in the temperature range below the room temperature. Four selected spin structures are schematically illustrated in Fig. 1.

Fig. 1a shows the experimentally determined ground state magnetic structures [18, 19] (shortened as SDW-AFM). The Fe spins form an antiferromagnetic (AFM) collinear structure along the $a$-axis direction and ferromagnetic along the $b$-axis direction. The nearest-neighbor Fe spins along the $c$ axis are antiparallel in SDW-AFM. Fig. 1b shows the widely adopted magnetic structures (shortened as 2d-Stripe) in most of the existing theoretical calculations [13-17]. In fact, the 2d-Stripe structure is two-dimensional. Within the $ab$ plane, the magnetic structures of the 2d-Stripe structure are identical to that of SDW-AFM while the nearest-neighbor Fe spins along the $c$ axis are parallel. Fig. 1c and Fig. 1d show two additional low energy magnetic structures that we have labeled as "STR6398" and "STR6980".

To calculate the 0 K energies, we employed the projector-augmented wave (PAW) method [20, 21] within the generalized gradient approximation (GGA) of Perdew-Burke-Ernzerhof (PBE) [22] as implemented in the VASP package [20, 21]. Total supercell energies were converged to 10$^{-6}$ eV. A plane wave cutoff of 348.2479 eV and a 6×6×4 $\Gamma$- centered *k*-point mesh are used. For the lattice vibration, we find that the Debye-Grüneisen approach [23, 24] is a fast and yet accurate enough solution for the present topic. The calculation of $F_{el}^{\sigma}$ in Eq. (5) follows the previous work [11].

Figure 2 presents the first-principles 0 K supercell total energies of the 13 non-equivalent spin structures as a function of supercell volume. It is seen that the ground state is SDW-AFM, in agree with the measurement [18, 19]. We find that the first excited state is STR6398 and the exciting energy is ~26.4 meV. It is also noted that from Fig. 2 the energies of most of spin structures studied in this Letter are lower than that of 2d-Stripe structure which are widely adopted as the ground state in the literatures [13-17]. In fact, we will also see in Fig. 3 that the thermal population of the 2d-Stripe structure is quite small. This raises a doubt about all the existing theoretical models based on the 2d-Stripe structure. Therefore, the magnetism in iron pnictides is more appropriately categorized as anisotropic three-dimensional [25, 26].

Figure 3 depicts calculated thermal populations ($x^{\sigma}$) of the 13 spin structures as a function of temperature. For comparison, the experimentally measured (1,0,3) magnetic peak intensity [27, 28] of the SDW-AFM structure as a function of temperature are also shown. Our theory predicts very well the measured data by Huang et al. [27]. From Fig. 3, it is seen that the thermal population of the SDW-AFM structure decreases fast above 50 K. This indicates the starting of the magnetic phase transitions. To clearly show the transition, we have calculated the specific heat. In fact, the magnetic specific heat due to the configurational coupling from Eq. (3) is

$$C_m = \frac{1}{k_B T^2} \left\{ \sum_\sigma x^\sigma (E^\sigma)^2 - \left[ \sum_\sigma x^\sigma E^\sigma \right]^2 \right\}, \tag{6}$$

with the internal energy given by $E^\sigma = F^\sigma + TS^\sigma$ and the entropy of electronic state $\sigma$ given by $S^\sigma = -(\partial F^\sigma / \partial T)_V$. Equation (6) is a generalization of the Schottky specific heat anomaly for a two-state system [10]. Figure 4 shows our predicted temperature evolution of the magnetic specific heat. For comparison, we also plot the measured magnetic specific heat that we extracted from the measured total specific heat by Ni et al. [8] and the lattice specific heat by Mittal et al. [7] using the lattice dynamics approach in fitting the measured phonon density-of-states. We clearly see a peak along the specific heat curve at ~100 K which is Schottky anomaly. We want to add here that, at the present, our model has not considered the effect of the long range magnetic field resulted from long range ordering. Since our system is limited to a system of 40 atoms, the slightly Sn-doped system of Ni et al. [8] is possibly more appropriate to demonstrate our theory. In fact, we note that the measured SDW ordering temperatures are scattered in the range of 80 – 150 K [7-9, 27, 28].

Figure 5 shows the calculated SDW ordering temperature ($T_{SDW}$) as a function of temperature. We have determined $T_{SDW}$ at which the thermal population of the SDW-AFM spin structure equals to 50%. In comparison, the measured $T_{SDW}$ by Fukazawa et al. [29] are also plotted in Fig. 5. An attempt is made to calculate a characteristic temperature $T_*$ at which the thermal population of the SDW-AFM spin structure equals to 99.99%. Both show remarkable agreement with experimental data.

In summary, we presented a first-principles formulation that incorporates coupling between spin fluctuations and lattice vibrations for BaFe$_2$As$_2$. We quantitatively predicted the pressure dependence of the spin density wave ordering, the Schottky anomaly in the specific heat, and the temperature dependence of the thermal populations of spin structures. Our approach is applicable to many other highly correlated magnetic systems.

**Acknowledgements** Calculations were conducted at the LION clusters at the Pennsylvania State University, supported in part by NSF Grants Nos. DMR-9983532, DMR-0122638, DMR-0205232, and DMR-0510180. This research also used resources of the NERSC supported by the Office of Science of the U.S. Department of Energy under the Contract No. DE-AC02-05CH11231. This work was supported in part by a grant of HPC resources from the Arctic Region Supercomputing Center at the University of Alaska Fairbanks as part of the Department of Defense High Performance Computing Modernization Program. The authors are indebted to Dr. R. Mittal for sending us the heat capacity data. Communications with Dr. N. Ni and Dr. W. Hermes are acknowledged.

**Figure Captions**

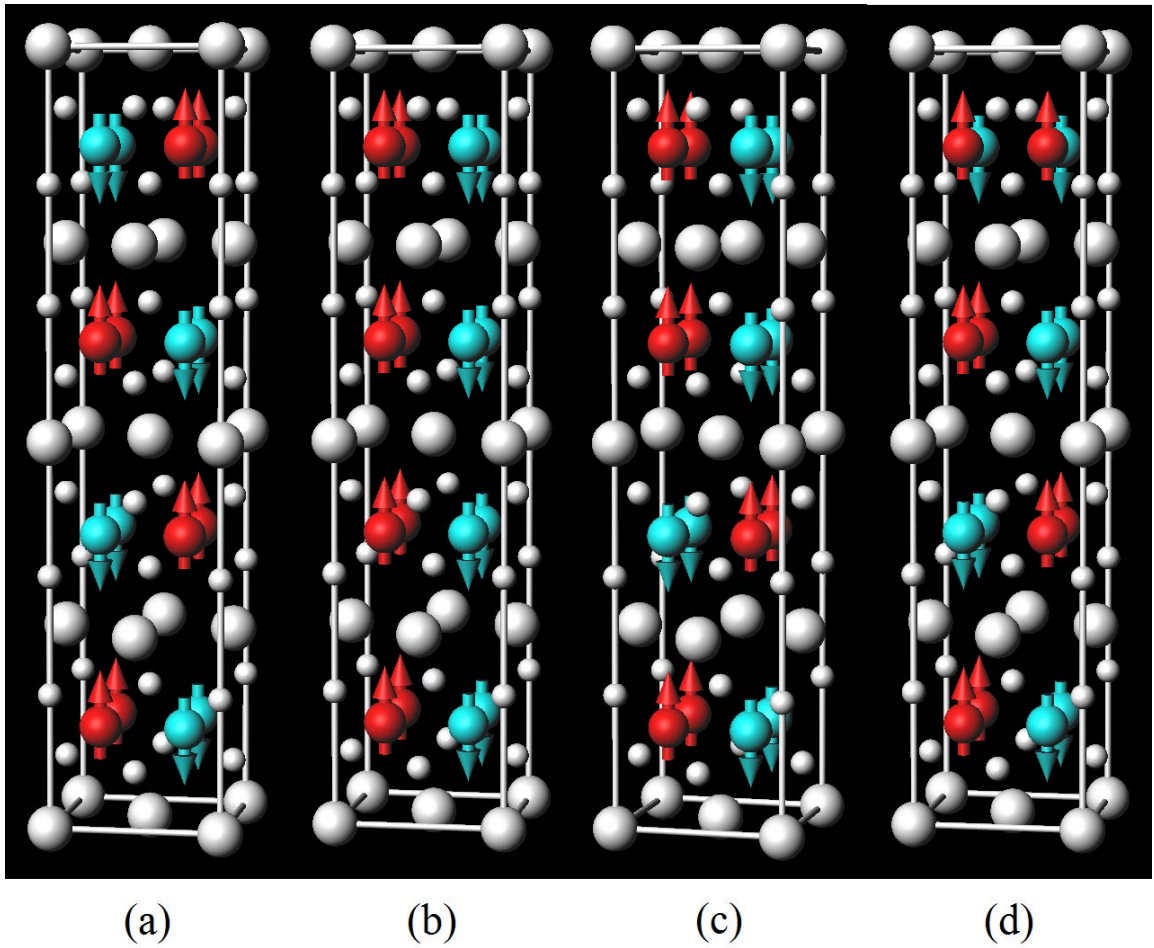

Figure 1. Selected spin structures of BaFe$_2$As$_3$ with (a) SDW-AFM; (b) 2d-Stripe; (c) STR6398; and (d) STR6980. Big grey balls: Ba; Small grey balls: As; Red balls with up arrows: Fe with spin up; Red balls with down arrows: Fe with spin down.

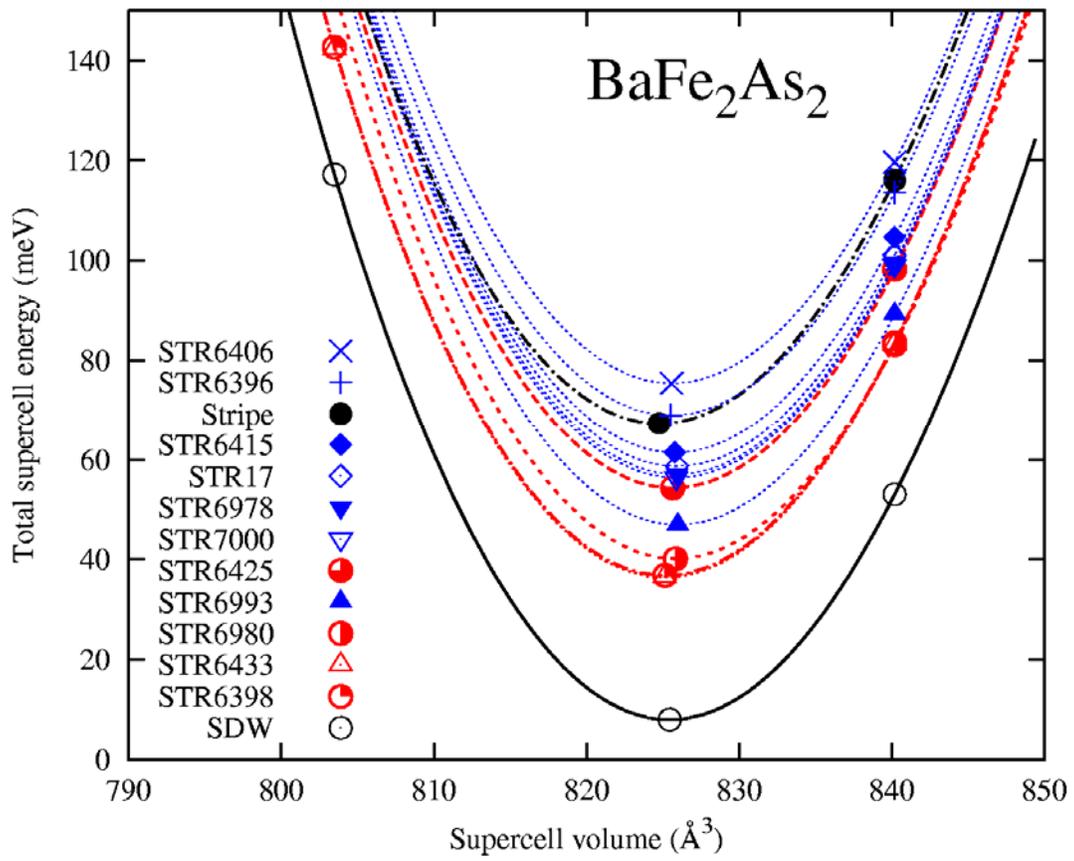

Figure 2. Calculated supercell energies of BaFe$_2$As$_2$ as a function of the $\sqrt{2}a \times \sqrt{2}a \times 4c$ 40-atom supercell volume.

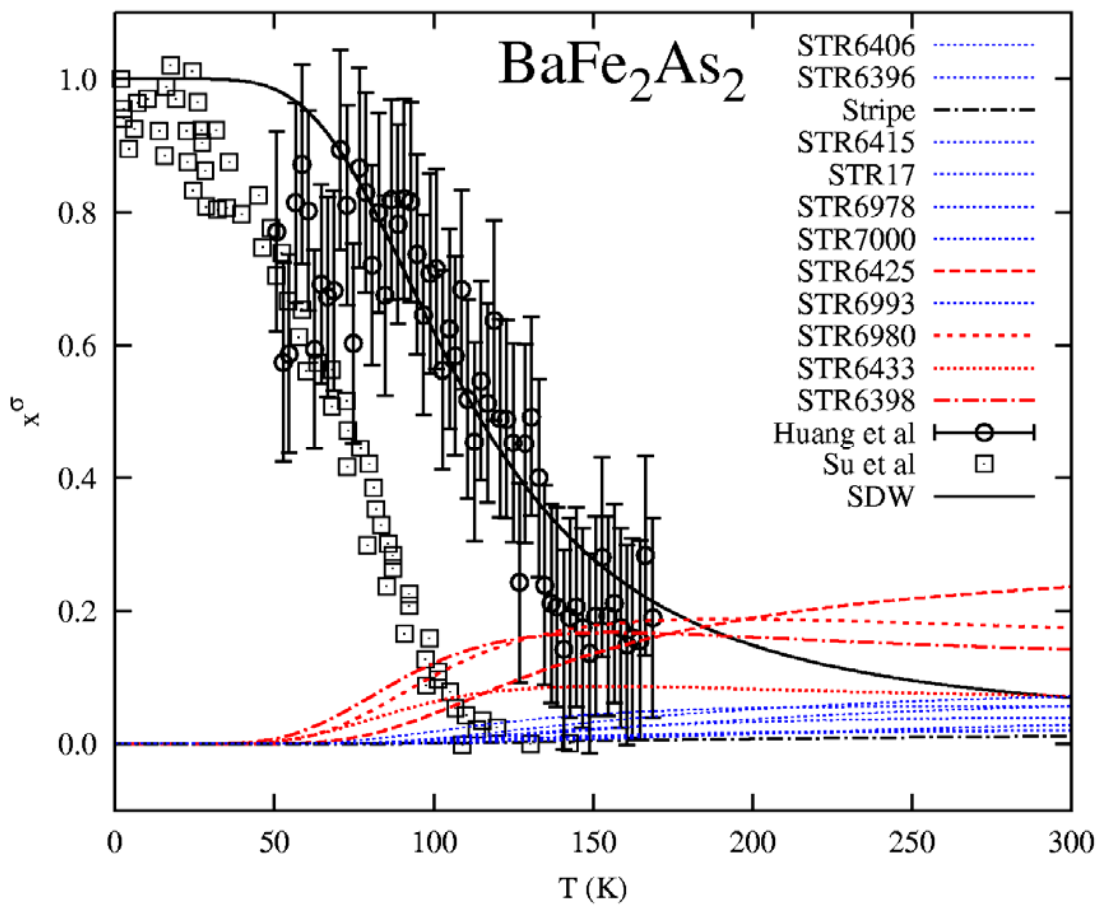

Figure 3. Thermal populations of the 13 spin structures of BaFe$_2$As$_2$ as a function of temperature. The lines represent the calculations. The circles with error bars are the measured (1,0,3) magnetic peak intensity of the SDW-AFM structure by Huang et al. [27]. The squares are the measured (1,0,3) magnetic peak intensity of the SDW-AFM structure by Su et al. [28].

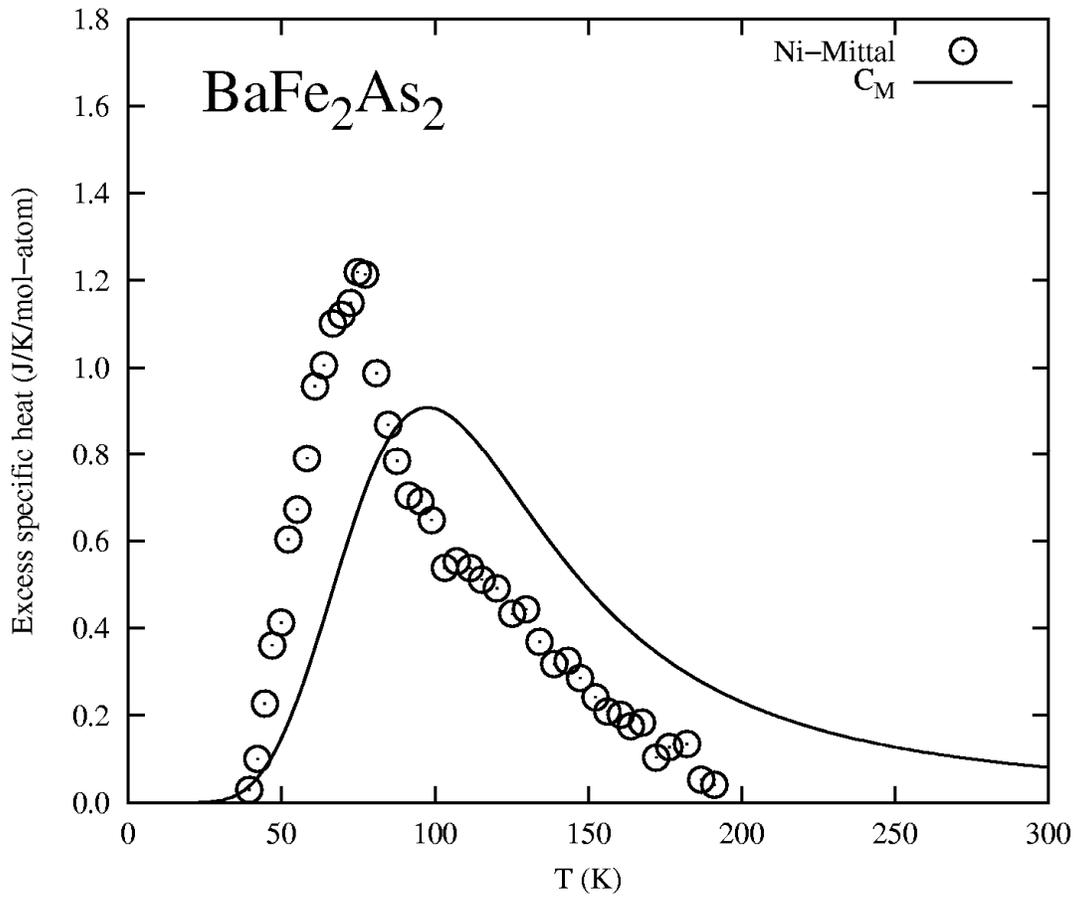

Figure 4. Magnetic specific heat ($C_m$, black line) of $BaFe_2As_2$. The line illustrates the calculated value. The open circles are the measured magnetic specific heat that we extracted from the measured total specific heat by Ni et al. [8] and the lattice specific heat by Mittal et al. [7] using the lattice dynamics approach in fitting the measured phonon density-of-states.

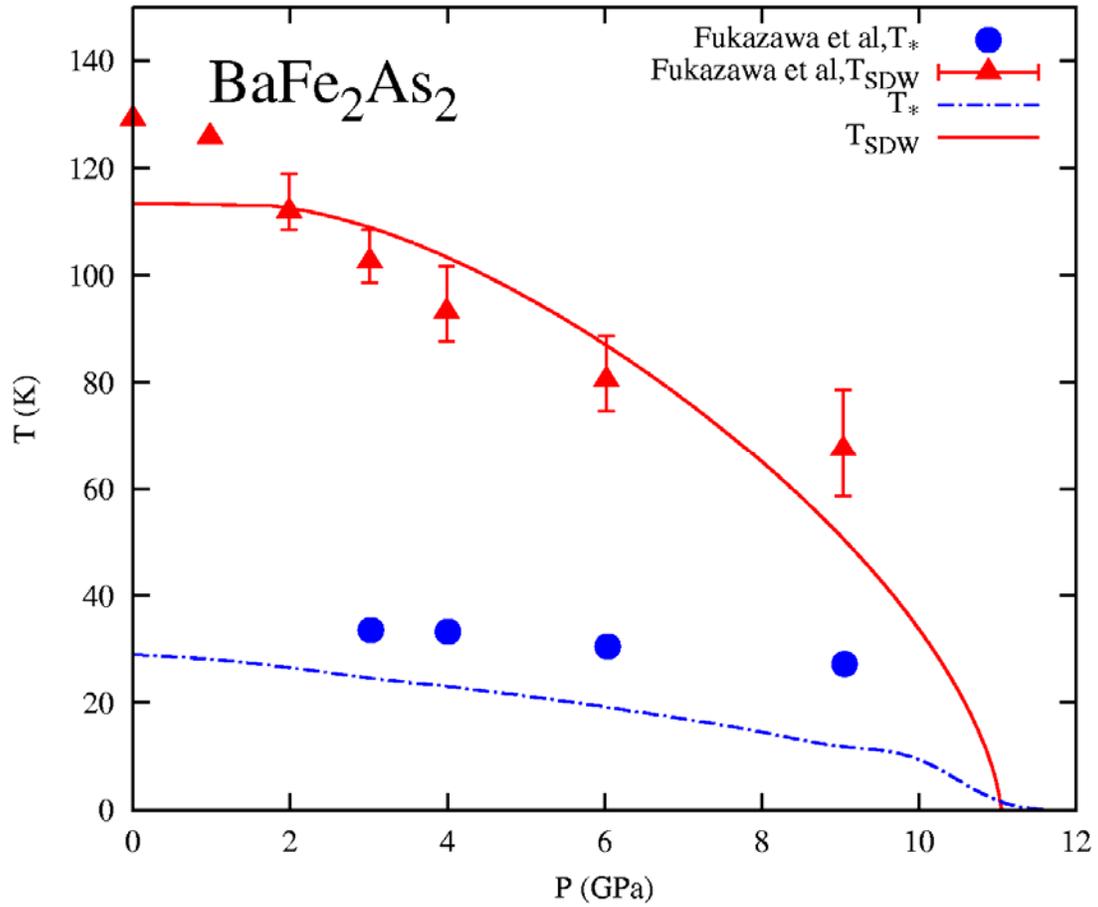

Figure 5. SDW ordering of BaFe$_2$As$_2$. The calculated ordering temperature $T_{SDW}$ and the characteristic temperature $T_*$ are shown by red solid line and blue dot-dashed line, respectively. The measured $T_{SDW}$ and $T_*$ by Fukazawa et al. [29] are shown by red triangles with error bars and blue circles, respectively